\makeatletter
\declare@file@substitution{revtex4-1.cls}{revtex4-2.cls}
\makeatother

\documentclass[two column]{aastex631}

\usepackage{graphicx}

\usepackage{natbib}
\usepackage{xcolor}
\usepackage{amsbsy,amsmath}

\begin{document}

\title{The Relation Between Variances of a 3D Density and Its 2D Column Density Revisited}

\author[0000-0003-4659-0758]{Heesun Yoon }
\affiliation{Department of Physics, College of Natural Sciences UNIST, Ulsan 44919, Republic of Korea}
\affiliation{Korea Astronomy and Space Science Institute, Daejeon, 34055, Republic of Korea}
\email{hsyoon@kasi.re.kr}

\author[0000-0003-1725-4376]{Jungyeon Cho}
\affiliation{Department of Astronomy and Space Science, Chungnam National University, Daejeon, 34134, Republic of Korea}
\correspondingauthor{Jungyeon Cho}
\email{jcho@cnu.ac.kr}

\begin{abstract}

We revisit the relation between the variance of three-dimensional (3D) density ($\sigma^{2}_{\rho}$) and that of the projected two-dimensional (2D) column density ($\sigma^{2}_{\Sigma}$) in turbulent media, which is of great importance in obtaining turbulence properties from observations.
Earlier studies showed that $\sigma^{2}_{\Sigma / \Sigma_{0}}/\sigma^{2}_{\rho / \rho_{0}} = \mathcal{R}$, where $\Sigma/\Sigma_0$ and $\rho/\rho_0$ are 2D column and 3D volume densities normalized by their mean values, respectively.
The factor $\mathcal{R}$ depends only on
the density spectrum for isotropic turbulence in a cloud that has similar dimensions along and perpendicular to the line of sight.
Our major findings in this paper are as follows.
First, we show that the factor $\mathcal{R}$ can be expressed in terms of $N$, the number of independent eddies along the line of sight. 
To be specific,  $\sigma^{2}_{\Sigma / \Sigma_{0}}/\sigma^{2}_{\rho/\rho_{0}}$ is proportional to  $\sim 1/N$, due to the averaging effect arising from independent eddies along the line of sight.
Second, we show that the factor $\mathcal{R}$ needs to be modified if the dimension of the cloud in the line-of-sight direction is different from that in the perpendicular direction.
However, if we express $\sigma^{2}_{\Sigma / \Sigma_{0}}/\sigma^{2}_{\rho / \rho_{0}}$ in terms of $N$, the expression remains same even in the case the cloud has different dimensions along and perpendicular to the line of sight. 
Third, when we plot $N\sigma^{2}_{\Sigma / \Sigma_{0}}$ against $\sigma^{2}_{\rho / \rho_{0}}$, two quantities roughly lie on a single curve regardless of the sonic Mach number, 
which implies that we can directly obtain the latter from the former.
We discuss observational implications of our findings.

\end{abstract}


\section{Introduction} \label{sec:intro}

The interstellar medium (ISM) is magnetized and highly turbulent \citep[see, for example,][]{Elmegreen2004}. Turbulence in the ISM affects many astrophysical processes, such as formation of dense cores and stars \citep[see][for a review]{Crutcher2012}, amplification of magnetic fields \citep{Meneguzzi1981, Cho2000, Schekochihin2004}, and cosmic ray accelerations \citep{Gaches2021, Lemoine2022}. Therefore, study of turbulence is essential for understanding these astrophysical processes. Observations of column density, centroid velocity, and channel maps are commonly used to study turbulence in the ISM. In this paper, we focus on the relation
between
the variance of the three-dimensional (3D) density ($\sigma^{2}_{\rho / \rho_{0}}$) and that of the projected two-dimensional (2D) column density ($\sigma^{2}_{\Sigma / \Sigma_{0}}$) in turbulent media. Here $\rho$, $\rho_0$, $\Sigma$, and $\Sigma_0$ are the 3D density, the mean 3D density, the column density, and the mean column density, respectively.

The 3D-density variance is one of the most important quantities that characterize turbulence.
Earlier numerical and analytic studies discussed the relation between the 3D density and turbulence parameters, such as the sonic Mach number ($M_s$) or driving modes \citep{Padoan1997, Passot1998, Beetz2008,  Federrath2008, Federrath2010, Price2011, Konstandin_2012}.
In general, the relation between the variance of the 3D density and the sonic Mach number takes the form
\begin{equation}
\sigma^{2}_{\rho / \rho_{0}} = b^{2} M_{s}^2,
\label{eq:3Dms}
\end{equation}  
where $b$, a constant of order unity, depends on the driving mode of turbulence: 
$b=1/3$ for solenoidally driven turbulence and $b=1$ for compressively driven turbulence \citep{Federrath2008, Federrath2010}. 
Therefore, in order to constrain the value of $M_s$ and/or the driving mode of turbulence,
it is necessary to know the variance of the 3D density.

However, we cannot directly observe the 3D density. We can observe only the 2D column density projected on the plane of the sky.
Earlier studies discussed the relation between the 2D and the 3D-density variances. 
If we know the relation between them, we can obtain $M_{s}$ from the 2D observations. 
Indeed there are such studies (\citealt{Brunt2010, Burkhart2012}; see also \citealt{Cho2009} for a discussion in a different context).
\cite{Brunt2010} suggested a method of reconstructing the variance of the 3D density ($\sigma^{2}_{\rho / \rho_{0}}$) from observed 2D column density spectrum:
\begin{equation}
 \frac{ \sigma^2_{\Sigma/\Sigma_0} }{ \sigma^2_{\rho/\rho_0} } =\mathcal{R}.
\end{equation}
The Brunt factor $\mathcal{R}$ can be directly calculated from the column density spectrum, if the 
 line-of-sight thickness of the cloud is the same as the size of the cloud
projected on the plane of the sky.
On the other hand, using numerical simulations in which the number of independent eddies along the line of sight ($N$)
is small (i.e., $N\lesssim 2.5$),  \cite{Burkhart2012} derived the relation:
\begin{equation}
     \sigma^{2}_{\Sigma/\Sigma_0}  = (b^{2}M_{s}^{2}+1)^{A} -1, 
     \label{eq:burkhart}
\end{equation}
where $b$ is a constant of order unity and $A$ is the scaling parameter.
They found the best-fit values of $b=1/3$ and $A=0.11$ for solenoidally driven turbulence.

We note that the number of independent eddies along the line of sight ($N$) affects
 the variance of the column density.
Suppose we have many independent eddies along the line of sight.
 Here we assume that all eddies have
similar density statistics. 
That is, we assume that they have similar mean 3D densities ($\sim \rho_0$) and 3D-density fluctuations ($\sim \sigma_\rho$).
 In this case, if we integrate the 3D density along the line of sight,
the fluctuating component will add up stochastically while
the mean component increases  linearly as the number of eddy increases.
As a result, if there are $N$ independent eddies along the line of sight, the standard deviation of
the normalized column density ($\sigma_{\Sigma/\Sigma_0}$) will decrease in proportion to $\sim 1/\sqrt{N}$.
Similarly, we can show that elongation of the cloud in the line-of-sight direction also affects the ratio 
$\sigma^2_{\Sigma/\Sigma_0} / \sigma^2_{\rho/\rho_0}$ (see Section \ref{sect:rect} and Section \ref{sect:obtain3D}).

In this paper, we revisit the relation between the variance of the 3D density ($\sigma^{2}_{\rho / \rho_{0}}$) and that of the projected 2D column density ($\sigma^{2}_{\Sigma / \Sigma_{0}}$) in turbulent media.
We show how the ratio  $\sigma^2_{\Sigma/\Sigma_0} / \sigma^2_{\rho/\rho_0}$ is related to the number of
independent eddies along the line of sight and extend the discussions in \cite{Brunt2010} to the case
the line-of-sight thickness of cloud ($L_{\rm los}$ in Figure \ref{fig:shape}) is different from the size of the cloud
projected on the plane of the sky ($L_{\bot}$ in Figure \ref{fig:shape}).
In Section \ref{sect:method}, we describe the numerical setup and theoretical considerations. 
In Section \ref{sect:result}, we present our major findings. 
In Section \ref{sect:dis},  we give discussions. 
In Section \ref{sect:summary}, we summarize our findings.

\section{Numerical Methods and Theoretical Considerations} \label{sect:method}

\subsection{Numerical Methods}

\subsubsection{Numerical Simulations} \label{sect:num}

\begin{deluxetable*}{ccccccc }
\tabletypesize{\scriptsize}
\tablecaption{Parameters of Simulations}
\tablewidth{0pt}
\tablehead{ 
\colhead{Model} & 
\colhead{Driving} & \colhead{Resolutions} & 
\colhead{B$_{0}/\sqrt{4 \pi \rho_0}$  \tablenotemark{a}} & 
\colhead{$M_{A}$  \tablenotemark{b}} &
\colhead{$M_{s}$  \tablenotemark{c}} & 
\colhead{$k_{f}$  \tablenotemark{d}}  
} 
\startdata
solF512-kf10b1  & Solenoidal   & $512^3$  & 1.0 & $\sim$ 0.93 $\pm$ 0.01 & $\sim$ 0.66  $\pm$ 0.01  
                               & 10   \\
                &              & $512^3$  & 1.0 & $\sim$ 0.88 $\pm$ 0.02 & $\sim$ 0.88  $\pm$ 0.02  
                               & 10    \\
                &              & $512^3$  & 1.0 & $\sim$ 0.94 $\pm$ 0.02 & $\sim$ 2.98  $\pm$ 0.02  
                               & 10    \\
                &              & $512^3$  & 1.0 & $\sim$ 1.06 $\pm$ 0.02 & $\sim$ 4.74  $\pm$ 0.02  
                               & 10    \\
                &              & $512^3$  & 1.0 & $\sim$ 0.86 $\pm$ 0.01 & $\sim$ 8.62  $\pm$ 0.01  
                               & 10   \\
\hline
solF1024-kf10b1 &  Solenoidal  & $1024^3$ & 1.0 & $\sim$ 1.00 $\pm$ 0.01 & $\sim$ 3.16  $\pm$ 0.01  
                               & 10   \\
                &              & $1024^3$ & 1.0 & $\sim$ 1.04 $\pm$ 0.01 & $\sim$ 10.40 $\pm$ 0.01  
                               & 10   \\
\hline
solF512-kf2.5b1 &  Solenoidal  & $512^3$  & 1.0 & $\sim$ 0.80 $\pm$ 0.02 & $\sim$ 0.81 $\pm$ 0.02  
                               & 2.5   \\
                &              & $512^3$  & 1.0 & $\sim$ 0.85 $\pm$ 0.02 & $\sim$ 3.40 $\pm$ 0.02  
                               & 2.5   \\
                &              & $512^3$  & 1.0 & $\sim$ 0.87 $\pm$ 0.02 & $\sim$ 5.24 $\pm$ 0.02  
                               & 2.5  \\
                &              & $512^3$  & 1.0 & $\sim$ 0.86 $\pm$ 0.03 & $\sim$ 6.87 $\pm$ 0.03 
                               & 2.5   \\
\hline \hline
compF512-kf2.5b1 & Compressive  & $512^3$  & 1.0 & $\sim$ 1.00 $\pm$ 0.02 & $\sim$ 0.50 $\pm$ 0.02   
                               & 2.5  \\
                &              & $512^3$  & 1.0 & $\sim$ 0.98 $\pm$ 0.02 & $\sim$ 0.98 $\pm$ 0.02   
                               & 2.5   \\
                &              & $512^3$  & 1.0 & $\sim$ 1.06  $\pm$ 0.02 & $\sim$ 3.34 $\pm$ 0.02  
                               & 2.5 \\
                &              & $512^3$  & 1.0 & $\sim$ 0.94 $\pm$ 0.01 & $\sim$ 9.20 $\pm$ 0.01   
                               & 2.5 \\
\hline \hline
solF512-kf2.5Hydro &   Solenoidal  & $512^3$  & 0.0 & $\infty$               & $\sim$ 1.02  $\pm$ 0.02  
                               & 2.5     \\
                &              & $512^3$  & 0.0 & $\infty$               & $\sim$ 10.25 $\pm$ 0.01  
                               & 2.5    \\
\enddata
\tablecomments{
\tablenotetext{a}{The Alfv\'{e}n speed of the uniform magnetic field.} 
\tablenotetext{b}{The Alfv\'{e}n Mach number. $M_{A}$ is defined by $M_{A} = v_{\rm rms}/v_{A}$, where $v_{\rm rms}$ is the rms velocity
    and  $v_{A}$ 
     is the Alfv\'{e}n speed.}
\tablenotetext{c}{The Sonic Mach number. $M_{s}$ is defined by $M_{s} = v_{\rm rms}/c_{s}$, where $c_{s}$ is the isothermal sound speed. }
\tablenotetext{d}{The driving wave number, which is similar the number of eddies along the line of sight ($N$). }
}
\label{tbl:sim}
\end{deluxetable*}

We solve the ideal isothermal magnetohydrodynamic (MHD) equations in a periodic computational box of size $2\pi$ with either $512^3$ or $1024^3$ grid points:

\begin{eqnarray}
\frac{\partial \rho}{\partial t} + \nabla \cdot (\rho \textbf{\textit {v}})= 0, \\
\rho(\frac{\partial \textbf{\textit{v}}}{\partial t} + \textbf{\textit{v}} \cdot \nabla \textbf{\textit{v}}) + \nabla p -\frac{1}{4 \pi} (\nabla \times \textbf{\textit{B}}) \times \textbf{\textit{B}} = \rho \textbf{\textit{f}},  \label{eq:v} \\
\frac{\partial \textbf{\textit{B}}}{\partial t} - \nabla \times (\textbf{\textit{v}} \times \textbf{\textit{B}}) = 0,   \label{eq:b} \\
\nabla \cdot \textbf{\textit{B}} = 0,   \label{eq:div}
\label{eq:MHD}
\end{eqnarray} 
 where  $\rho$ is the gas density,  $\bold{v}$ is the velocity, $p = c_{s}^2\rho$, $p$ is the pressure, $c_{s}$ is the isothermal sound speed, $\bold{B}$ is the magnetic field, and $\bold{f}$ is the driving force. 
 In all simulations,  the density at  $t=0$  is set to unity (i.e., $\rho=\rho_0=1$ at $t=0$) and the rms velocity $v_{\rm rms}$ after saturation  is between $\sim 0.8$ and $\sim 1.0$.
The magnetic field consists of a uniform magnetic field ($\bold{B_0}$) and a fluctuating magnetic field ($\bold{b}$): $\bold{B} = \bold{B_0} + \bold{b}$. The uniform magnetic field $\bold{B_0}$ is placed along the $x$-axis and 
 the  Alfv\'{e}n velocity of the uniform field is unity: $v_{A} = B_0 / \sqrt{4 \pi \rho_{0}}=1$. Therefore, the Alfv\'{e}n Mach number ($M_{A} \equiv v_{\rm rms} / v_{A}$) is between $\sim 0.8$ and $\sim 1.0$, which means that we focus on marginally sub-Alfv\'{e}n turbulence in this paper (see Table \ref{tbl:sim}). For comparison, we additionally perform hydrodynamic turbulence simulations. To do this, we set $\bold{B}=0$ in Equation (\ref{eq:v}) and do not solve Equations (\ref{eq:b}) and (\ref{eq:div}).

We drive turbulence by applying two types of driving: solenoidal ($\nabla \cdot \bold{f} = 0$) and compressive ($\nabla \times \bold{f}= 0 $) drivings. In the case of solenoidal driving, we drive turbulence either on large scales ($k_{f} \sim 2.5 $ in Fourier space, where $k_f$ is the driving wave wavenumber) or on small scales ($k_{f} \sim 10$). In the case of compressive driving, we drive turbulence only on large scales ($k_{f} \sim 2.5 $).  
The driving scale ($L_{f}$) is equal to $\sim L/k_{f}$, where $L$ is the size of the computational box ($=2\pi$ in our simulations). The number of independent eddies along the line of sight is $ N \sim L/L_{f} \sim k_{f}$ \citep{Cho2016, Yoon2019}.
Therefore, if the length of the line of sight is equal to $L$, $k_f \sim 2.5$ and $\sim 10$ correspond to $N\sim 2.5$ and $\sim 10$, respectively. 

We generate the 3D turbulence data for various sonic Mach numbers ($M_{s}$), which  is the ratio of the rms velocity ($v_{\rm rms}$) to the sound speed ($c_{s}$):  $M_{s} = v_{\rm rms} / c_{s}$. Table \ref{tbl:sim} summarizes the values of $M_{s}$, $M_{A}$, and $k_f$ we use in this paper.

\subsection{Theoretical considerations: the relation between $\sigma^{2}_{\Sigma / \Sigma_{0}}$
and $\sigma^{2}_{\rho / \rho_{0}}$ }  \label{sect:theory}
\begin{figure*}
\centering
\includegraphics[scale=.20]{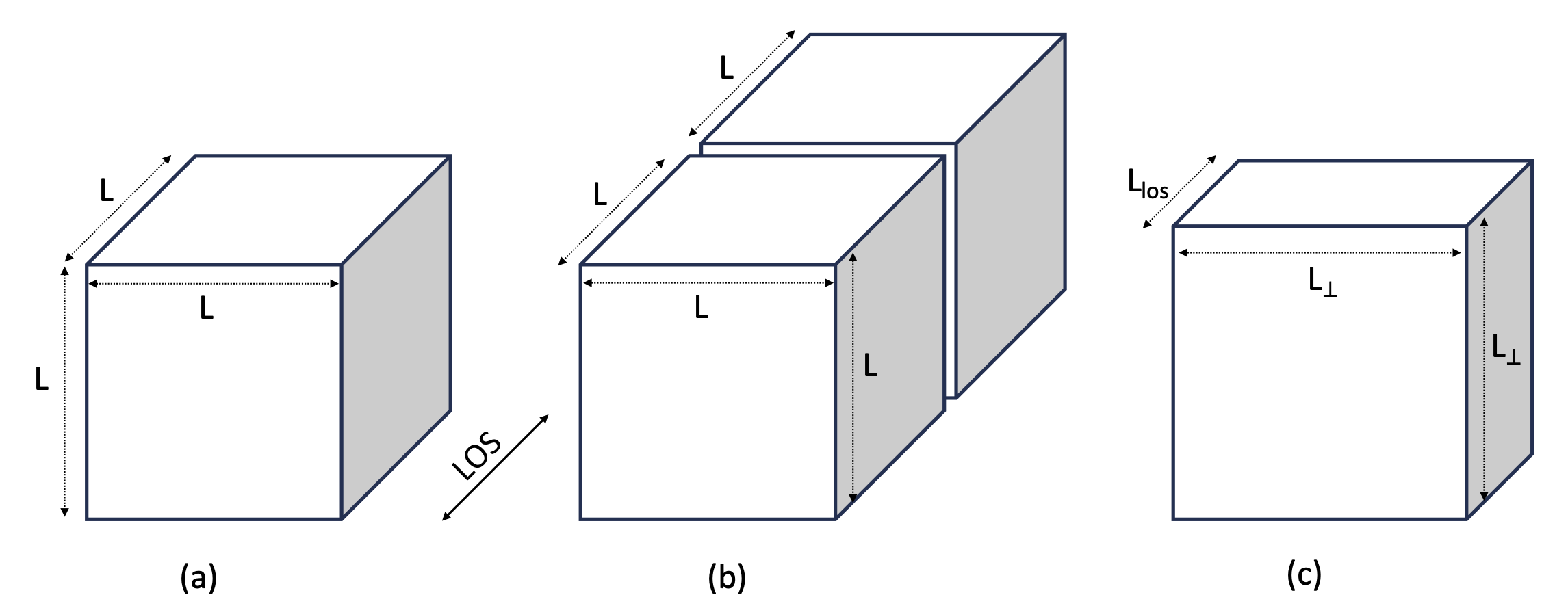} 
\caption{Cloud shapes. (a) A cubic cloud. (b) Two cubic clouds. (c) A rectangular cloud with dimension $L_\bot \times L_\bot \times L_{\rm los}$. }
\label{fig:shape}
\end{figure*}

In this subsection, we rederive the relation between the 3D-density and the projected 2D-density fluctuations.
For simplicity, we consider idealized cloud shapes: a cubic and a rectangular (to be exact, a rectangular cuboid) clouds. 
Throughout the paper (except Section \ref{sect:small}), we assume that the driving scale of turbulence is smaller than the size of the cloud projected on the plane of the sky (i.e., $L_f < L$ or $L_f < L_\bot$; see Figure \ref{fig:shape}).
We discuss the relation in a cubic cloud first and then generalize it to a rectangular one.
The relation derived for a  cubic cloud is virtually identical to the one in \cite{Brunt2010}.

\subsubsection{A cubic cloud with dimension $L \times L \times L$}

The observed column density $\Sigma$ is the density summed along the line of sight.
 If the Fourier transform of the 3D density $\rho$ is $\hat{\rho}(k_x,k_y,k_z)$,  the Fourier transform of $\Sigma$ is proportional to
$\hat{\rho}(k_x,k_y, 0)$, where we assume the line of sight is along the $z$-direction \citep[see, for example,][]{CL2002}.
To be exact,  the Fourier transform of $\Sigma$ is 
\begin{equation}
   L  \hat{\rho}(k_x,k_y, 0),
\end{equation}
where $L$ is the length of the cloud along the line of sight (see Figure \ref{fig:shape}(a)). 
The variance of $\Sigma$ is
\begin{equation}
    \sigma^2_{\Sigma} =  L^2  \sum_{k_x, k_y}| \hat{\rho}(k_x, k_y, 0) |^2,
\end{equation}
where we exclude  $k_x=k_y=0$ in the summation.
On the other hand, the variance of the 3D density $\rho$ is
\begin{equation}
    \sigma^2_{\rho} =  \sum_{k_x, k_y, k_z}| \hat{\rho}(k_x, k_y, k_z) |^2,
\end{equation}
where we exclude  $k_x=k_y=k_z=0$ in the summation.
Therefore, we have
\begin{equation}
   \frac{ \sigma^2_{\Sigma/\Sigma_0} }{ \sigma^2_{\rho/\rho_0} }
   \equiv \frac{ \sigma^2_{\Sigma} /(L^2 \rho_0^2) }{ \sigma^2_{\rho}/\rho_0^2 }
   =  \frac{     \sum_{k_x, k_y}| \hat{\rho}(k_x, k_y, 0) |^2 }
            {\sum_{k_x, k_y, k_z}| \hat{\rho}(k_x, k_y, k_z) |^2 },
\end{equation}
where $\Sigma_0=L \rho_0$. 

If turbulence is isotropic, we have
\begin{equation}
   \frac{ \sigma^2_{\Sigma/\Sigma_0} }{ \sigma^2_{\rho/\rho_0} }
   =  \frac{     \sum_{ k \leq \sqrt{ k_x^2+k_y^2} < k+1}  2 \pi k |\hat{\rho}(k_x, k_y, 0) |^2 }
            {   \sum_{ k \leq \sqrt{ k_x^2+k_y^2} < k+1}  4 \pi k^2 | \hat{\rho}(k_x, k_y, 0) |^2  }
            \equiv \mathcal{R},
\end{equation}
where $\mathcal{R}$ is the `Brunt factor'.

If turbulence has a power-law spectrum
\begin{equation}
    |\hat{\rho}(k_x, k_y, 0) |^2  \propto \begin{cases}
             k^{-m}, & \text{ if } k \geq k_f ,\\
             0, & \text{ if } k < k_f ,
\end{cases}
\end{equation}
where $k_f$ corresponds to the wavenumber of the driving scale and we assume $m>3$, 
we have
\begin{equation}
     \frac{ \sigma^2_{\Sigma/\Sigma_0} }{ \sigma^2_{\rho/\rho_0} } = \mathcal{R}
       = \frac{ (3-m) }{2(2-m) } k_f^{-1} = \frac{ (3-m) }{2(2-m) } \frac{ L_f }{L }.
            \label{eq:cube}
\end{equation}
Since we assume that the length of the cloud along line of sight is also $L$, the ratio $L/L_f$ is the same as the number of independent eddies along the line of sight.
Therefore, we can write
\begin{equation}
     \frac{ \sigma^2_{\Sigma/\Sigma_0} }{ \sigma^2_{\rho/\rho_0} } 
        = \frac{ (3-m) }{2(2-m) } \frac{1}{N},   \label{eq:cubeN}
\end{equation}
where $N$ is the number of independent eddies along line of sight.
Dependence of the ratio on  $1/N$ in Equation (\ref{eq:cubeN}) is consistent with the expectation of the Central Limit Theorem.

\subsubsection{A rectangular cloud with dimension $L_\bot \times L_\bot \times L_{\rm los}$ }
\label{sect:rect}
Here we give a heuristic derivation of the relation between the 3D- and the projected 2D-density variances for a rectangular cloud. 
Suppose that we have two statistically identical cubic clouds along the line of sight (see Figure \ref{fig:shape}(b)).
Since two clouds are statistically identical, the shape of the observed column density spectrum will be similar to the one for a single cubic cloud. Note however that the amplitudes of the column density spectra for the rectangular and the cubic clouds are different: the Fourier amplitudes of the column density for the rectangular cloud will be $\sim \sqrt{2}$ times larger than those for the cubic cloud since each cloud contributes randomly. As a consequence, $\sigma_{\Sigma}^2$ for the former will be $\sim 2$ times larger than the one for the latter.  On the other hand, the mean column density $\Sigma_0$ for the rectangular cloud will be $\sim 2$ times larger than the one for the cubic cloud. Therefore, $\sigma_{\Sigma/\Sigma_0}^2$ will be $\sim 1/2$ times smaller than the one given in Equation (\ref{eq:cube}).

Let us extend the above discussion to the case of a rectangular box with dimension 
$L_\bot \times L_\bot \times L_{\rm los}$.
In this case, there are $L_{\rm los}/L_\bot$ cubes along the line of sight.
Therefore, $\sigma_{\Sigma/\Sigma_0}^2$ will become $L_{\rm los}/L_\bot$ times smaller than the one given in Equation (\ref{eq:cube}):
\begin{equation}
     \frac{ \sigma^2_{\Sigma/\Sigma_0} }{ \sigma^2_{\rho/\rho_0} } 
      =\mathcal{R} \frac{ L_\bot }{L_{\rm los} } .  \label{eq:rectR}
\end{equation}
 If the 3D power spectrum of density is proportional to $k^{-m}$ (with $m>3$), the above equation becomes
\begin{equation}
\begin{split}
     \frac{ \sigma^2_{\Sigma/\Sigma_0} }{ \sigma^2_{\rho/\rho_0} } 
        = \frac{ (3-m) }{2(2-m) } \frac{ L_f }{L_\bot } \frac{ L_\bot }{L_{\rm los} } \\
        =\frac{ (3-m) }{2(2-m) } \frac{ L_f }{L_{\rm los}}
        =\frac{ (3-m) }{2(2-m) } \frac{ 1}{N }.   \label{eq:rectN}
\end{split}
\end{equation}
Note that the last expression is identical to the one in Equation (\ref{eq:cubeN}).

In summary, we have showed in this subsection  that 
\begin{itemize}
\item{For a cubic cloud (with dimension $L \times L \times L$), $  \sigma^2_{\Sigma/\Sigma_0} / \sigma^2_{\rho/\rho_0}$ is equal to the Brunt factor $\mathcal{R}$,}

\item{For a rectangular cloud (with dimension $L_\bot \times L_\bot \times L_{\rm los}$), $  \sigma^2_{\Sigma/\Sigma_0} / \sigma^2_{\rho/\rho_0}$ is equal to  $\mathcal{R}\frac{ L_\bot }{L_{\rm los} }$, and}
\item{ 
$  \sigma^2_{\Sigma/\Sigma_0} / \sigma^2_{\rho/\rho_0} = \frac{ (3-m) }{2(2-m) } \frac{ 1}{N }$, 
where $N$ is the number of independent eddies along the line of sight,
regardless of the shape of the cloud.}
\end{itemize}

\subsection{Generating data with different $N$} \label{sect:N}
One of the main goals of this paper is to investigate the effect of $N$, the number of independent eddies
along the line of sight (see Equation (\ref{eq:cubeN}) or (\ref{eq:rectN})). 
To do this, we need to generate 3D data with different $N$.
We generate such data using two different methods.
\begin{itemize}
\item First, we generate \emph{cubic} data (see the shape in Figure \ref{fig:shape}(a)) using different driving scales (i.e., $k_f \sim 2.5$ and $k_f\sim 10$). In this case, we do not slice the data cubes. Instead, we use the whole data cubes. If we drive turbulence on different scales, we can naturally generate data cubes that have different number of independent eddies along the line of sight.  Note that the dimension of the observed cloud is $L \times L \times L$ and that $N \sim L/L_f \approx k_f$ in our simulations. 
\item Second, we generate \emph{rectangular} data  (see the shape in Figure \ref{fig:shape}(c)) as follows. We first generate a 3D data cube by driving turbulence at $k_f \sim 10$ and then divide a whole data cube into $n_{\rm slice}$ slices along the line of sight \citep[see][]{Yoon2019}. Since the original data cube has $\sim 10$ independent eddies along the line of sight, the average number of eddies along the line of sight in each slice is  $ N \sim 10/n_{\rm slice}$.
In this paper, we consider $n_{\rm slice} = 1, 2, 4$, and $8$, which corresponds to $N \sim 10, 5, 2.5$, and $1.25$. Note that the dimension of the observed cloud in this case is $L \times L \times (L/n_{\rm slice})$.
\end{itemize}

\section{Results} \label{sect:result}

\subsection{The variance of column density ($\sigma^{2}_{\Sigma / \Sigma_{0}}$) and the effects of $N$}
\label{sect:sigN}

\begin{figure*}
\centering
\includegraphics[scale=.33]{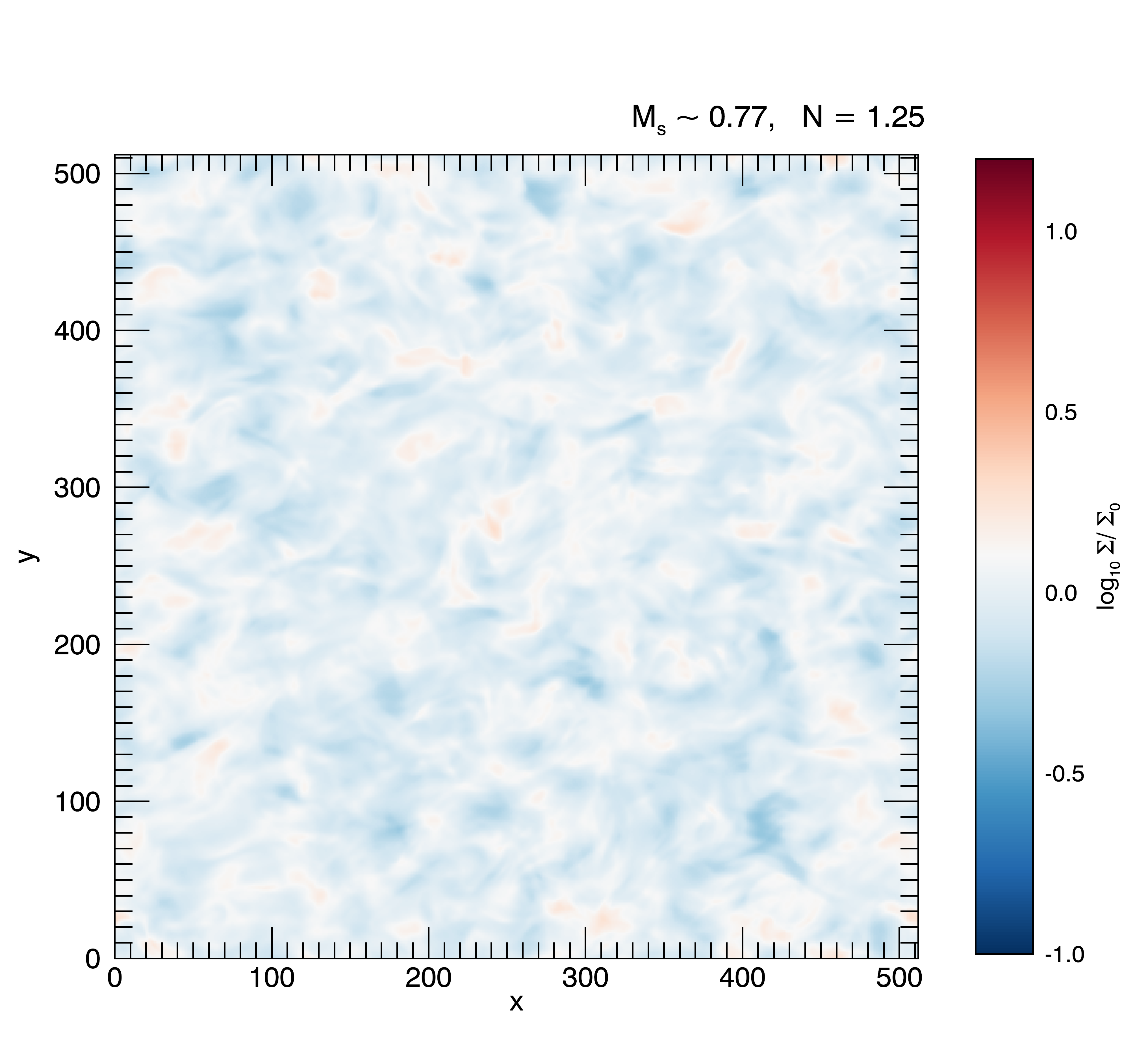}
\includegraphics[scale=.33]{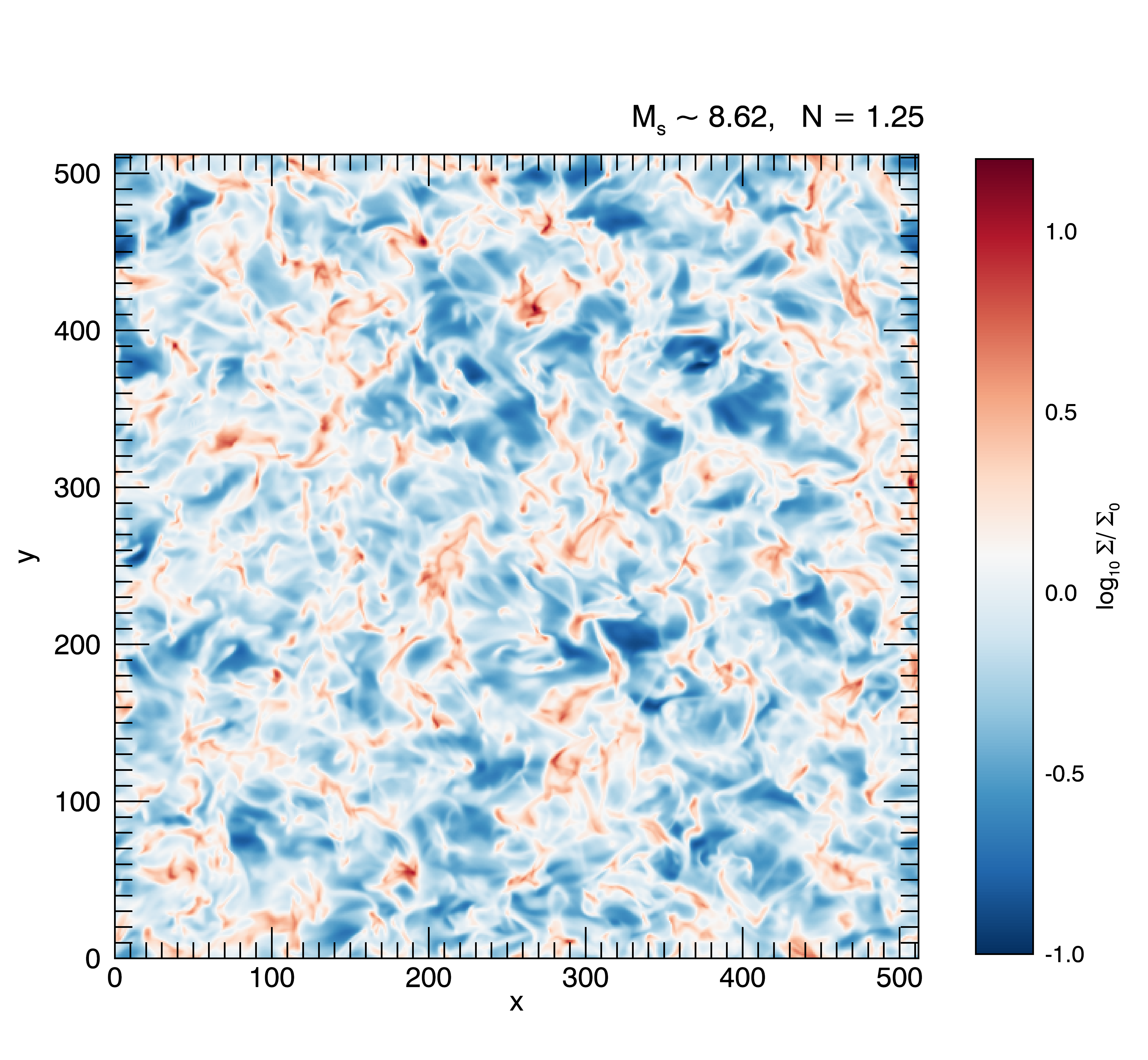} 
\includegraphics[scale=.33]{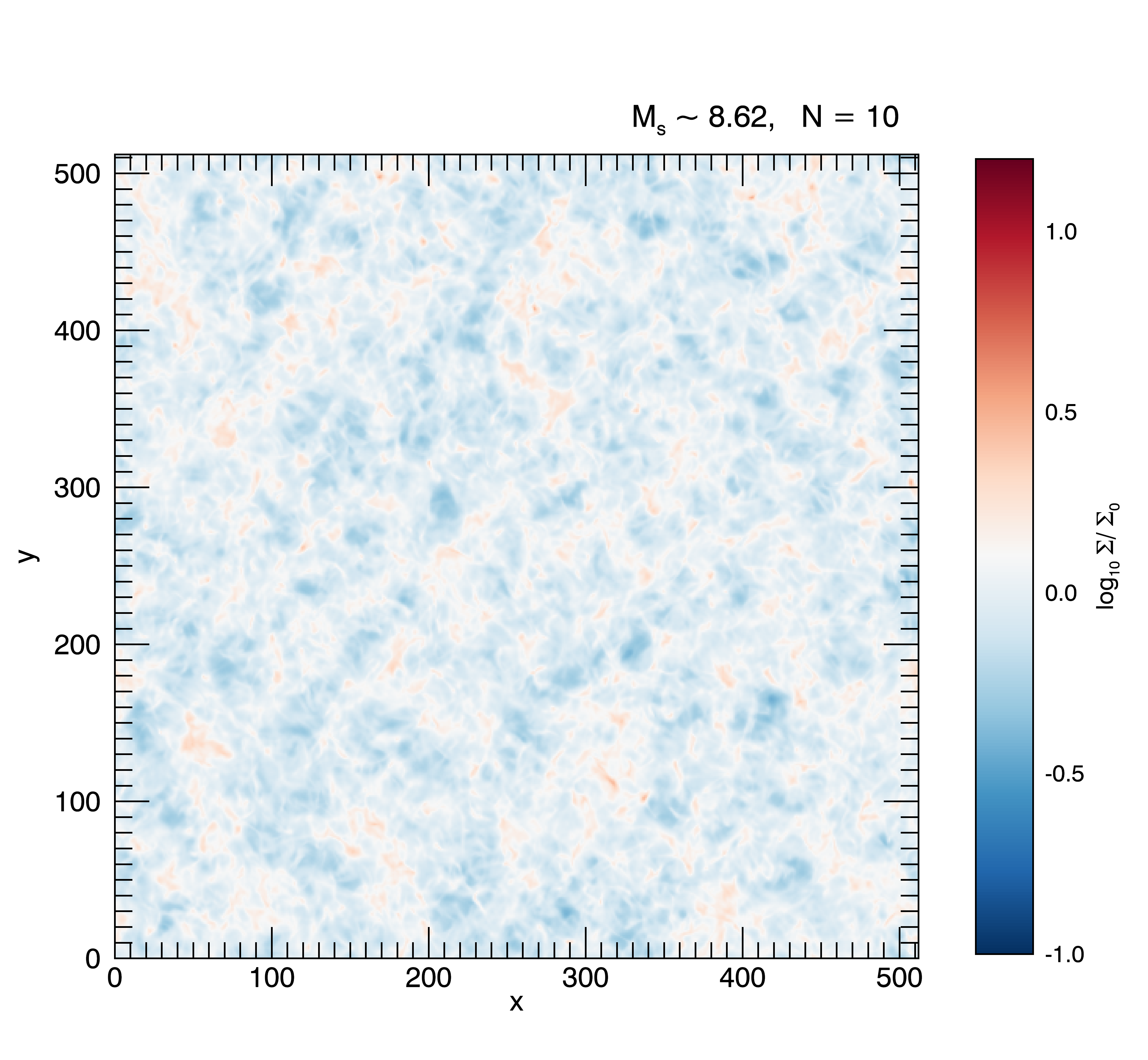} 
\caption{Normalized column density (i.e., $\log_{10}(\Sigma / \Sigma_{0})$) maps. The line of sight is along the $z$-axis, which is perpendicular to the mean magnetic field. (a) $M_{s} \sim 0.77$ and $N = 1.25$. (b) $M_{s} \sim 8.62$ and $N = 1.25$. (c) $M_{s} \sim 8.62$ and $N = 10$. }
\label{fig:vis}
\end{figure*}

In Figure \ref{fig:vis}, we visualize the normalized column density (i.e., $\log_{10}({\Sigma / \Sigma_{0}})$), in which the line of sight is perpendicular to the mean magnetic field. We use the same color bar range in all three panels. 
If we compare the left and the middle panels of  Figure \ref{fig:vis}, which have the same $N$ but different $M_s$, we can see that the column density map for smaller $M_s$ (the left panel) is smoother than that for larger $M_s$ (the middle panel).
If we compare the middle and the right panels of  Figure \ref{fig:vis}, which have the same $M_s$ but different $N$, we can see that the column density map for larger $N$ (the right panel) is smoother than that for smaller $N$ (the middle panel), which is due to the averaging effect.
Therefore it is important to note that both $M_s$ and $N$ affect the variance of the column density.

\begin{figure*}
\centering
\includegraphics[scale=.45]{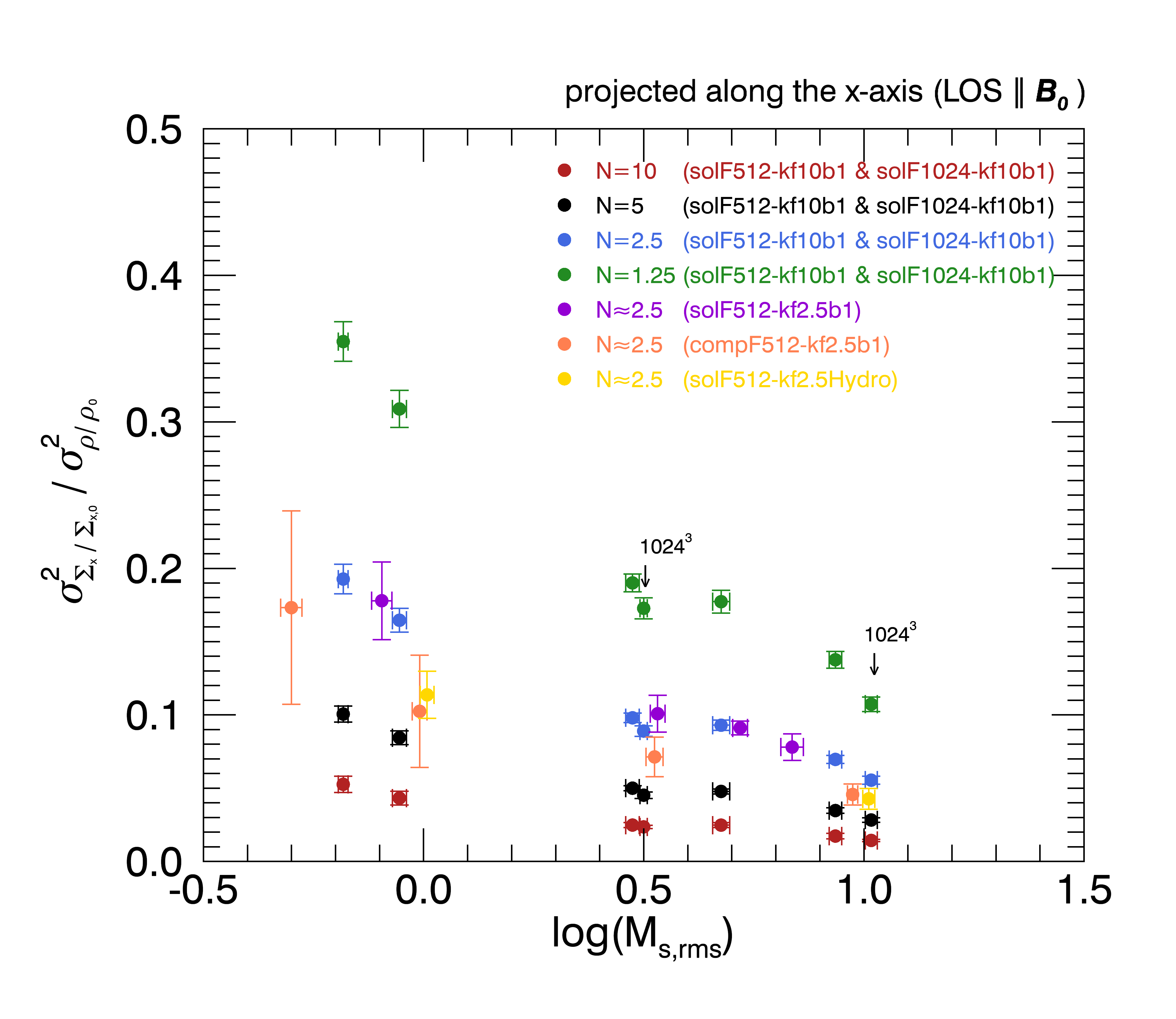}
\includegraphics[scale=.45]{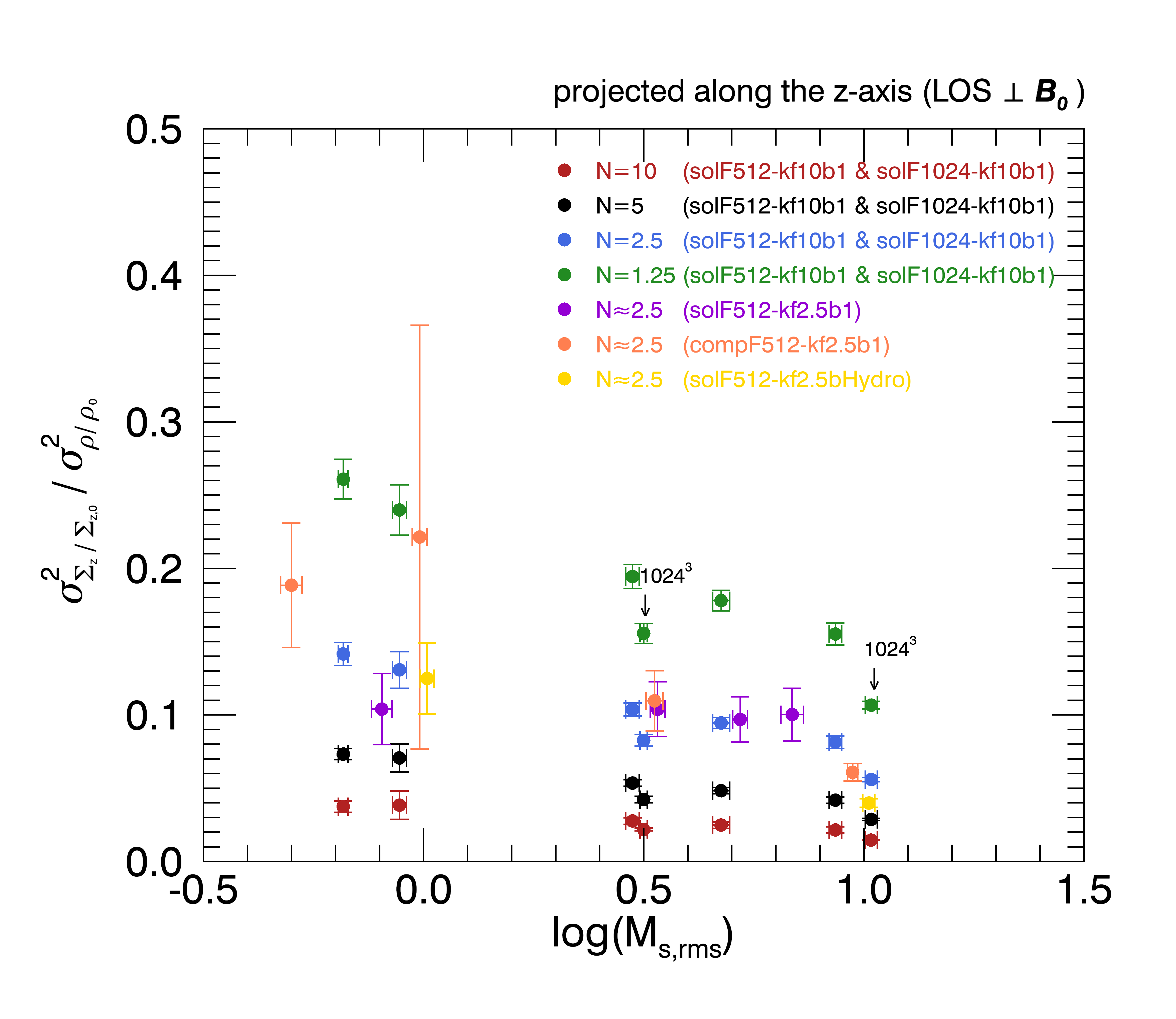} \\
\caption{Dependence of $ \sigma^2_{\Sigma/\Sigma_0} / \sigma^2_{\rho/\rho_0}$
on $M_s$ and $N$.
 The line of sight is (a) along the $x$-axis 
 and (b) along the $z$-axis. 
The red, black, blue, green circles are for $N$=10, 5, 2.5, and 1.25, respectively (see the second method in Section \ref{sect:N} for the technique of generating these data). 
Circles with other colors are for $N\approx 2.5$ (see the first method in Section \ref{sect:N} for the technique of generating the data). 
The purple and orange circles are for turbulence driven by large-scale ($k_{f} \sim 2.5$) solenoidal and compressive drivings, respectively. The yellow circles are for hydrodynamic turbulence driven by large-scale ($k_{f} \sim 2.5$) solenoidal driving.  The error bars represent one standard deviation of the column density fluctuation. 
}
\label{fig:R}
\end{figure*}

In Figure \ref{fig:R}, we plot $ \sigma^2_{\Sigma/\Sigma_0} / \sigma^2_{\rho/\rho_0}$ as function of the sonic Mach number $M_{s}$.
The left and the right panels are for the column densities projected along the $x$-axis (i.e., parallel to the mean magnetic field) and  the $z$-axis (i.e., perpendicular to the mean magnetic field), respectively.
We use the first method in Section \ref{sect:N} to generate data for  $N\approx 2.5$ and
the second method for $N = 10, 5, 2.5$, and $1.25$.
Data used for Figure \ref{fig:R} are listed  in Table \ref{tbl:obs}, which include both solenoidally and compressively driven turbulence.
As we can see in Figure \ref{fig:R}, $ \sigma^2_{\Sigma/\Sigma_0} / \sigma^2_{\rho/\rho_0}$ depends on both $M_{s}$ and $N$.
In particular, Figure \ref{fig:R} shows that the quantity  $ \sigma^2_{\Sigma/\Sigma_0} / \sigma^2_{\rho/\rho_0}$  is a decreasing function of $N$ (see also Table \ref{tbl:obs}).

\begin{figure*}
\centering  
\includegraphics[scale=.45]{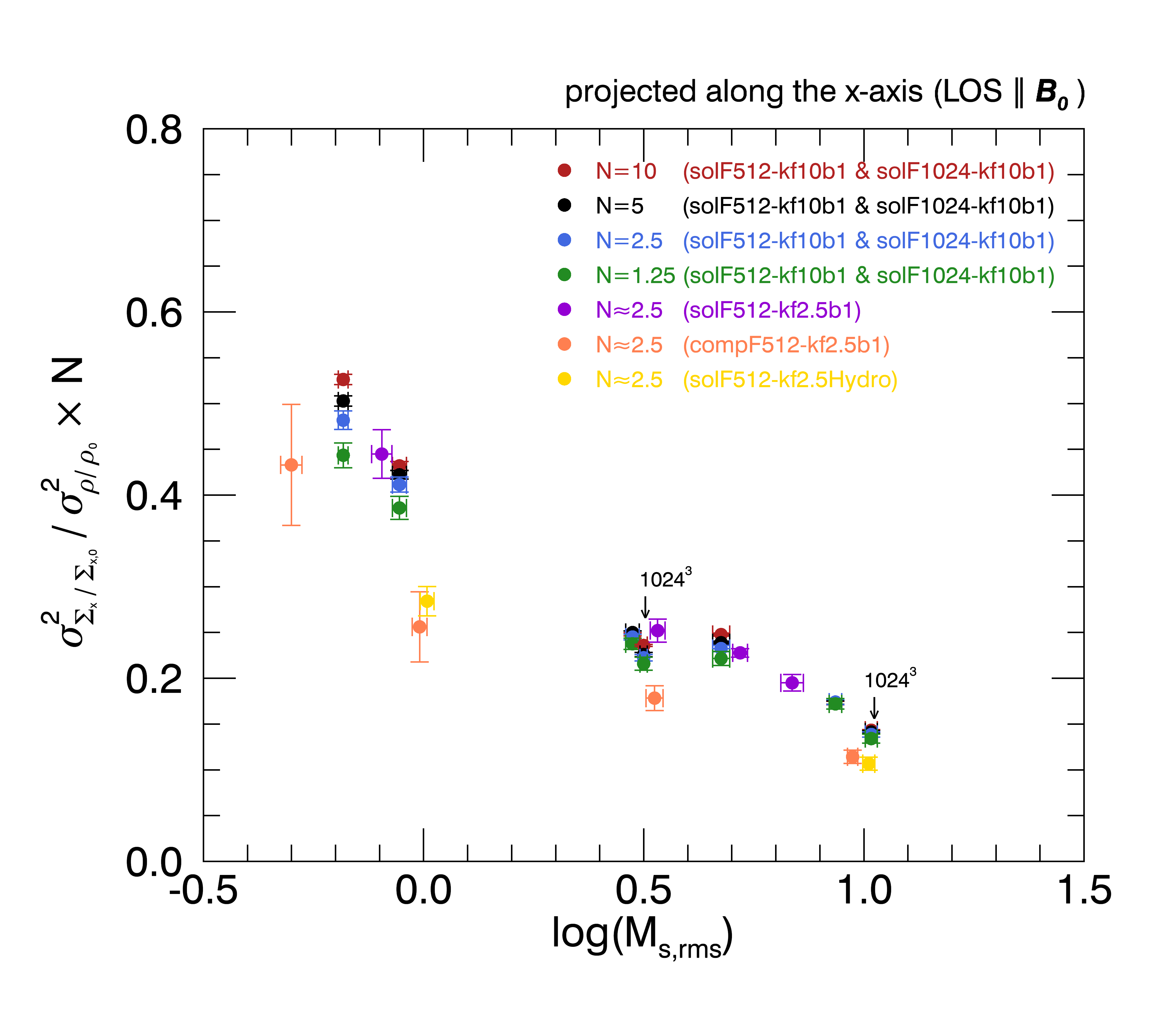}
\includegraphics[scale=.45]{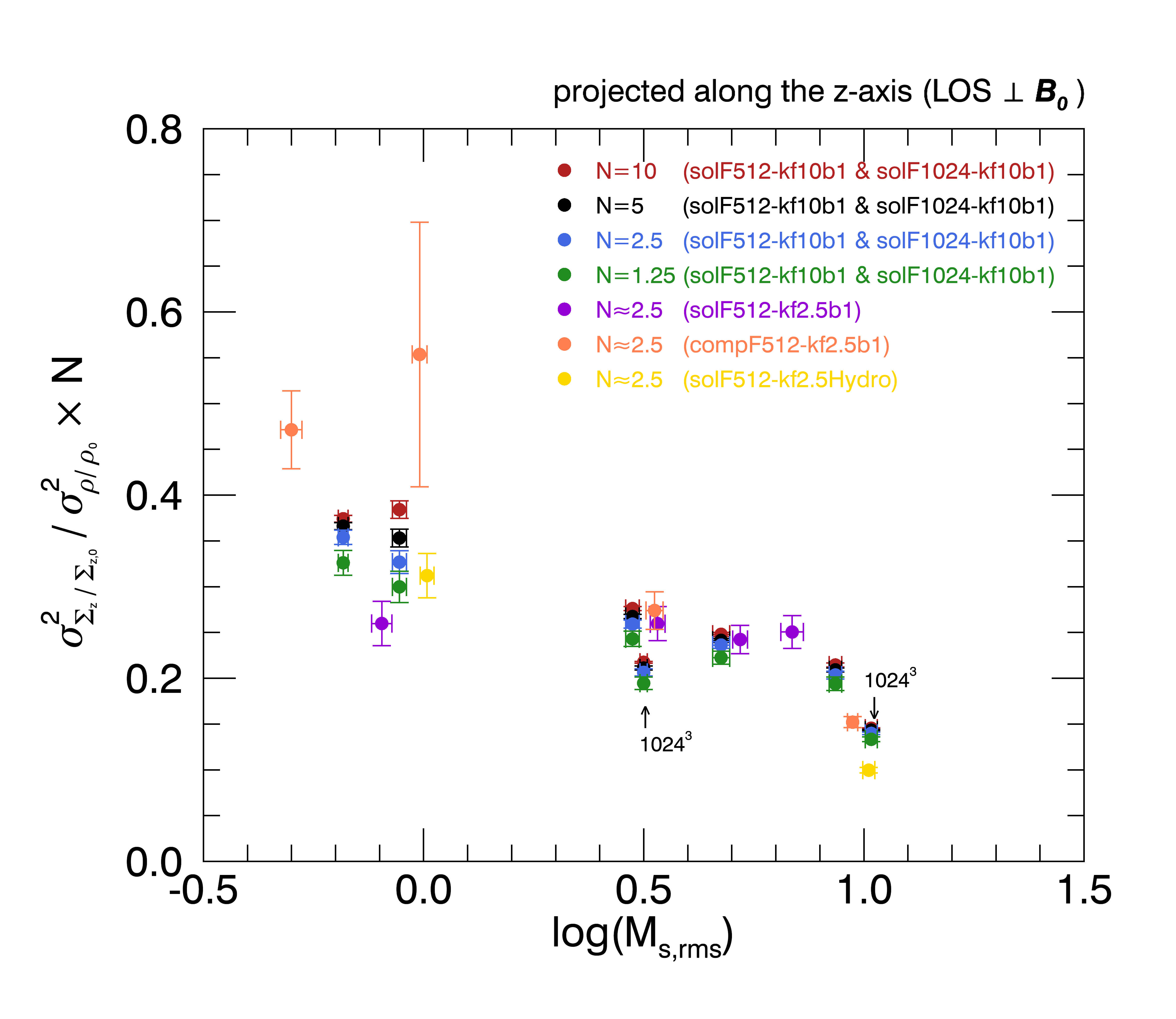}
\caption{
Dependence of $ \sigma^2_{\Sigma/\Sigma_0} / \sigma^2_{\rho/\rho_0}$ multiplied by $N$
 on $M_s$.
 The line of sight is (a) along the $x$-axis and (b) along the $z$-axis. The colors of the circles correspond to those in Figure \ref{fig:R}
}
\label{fig:RN}
\end{figure*}

Equations (\ref{eq:cubeN}) and (\ref{eq:rectN}) predict that  $ \sigma^2_{\Sigma/\Sigma_0} / \sigma^2_{\rho/\rho_0} \propto 1/N$.
Indeed, Figure \ref{fig:RN} shows that the quantity $ N \sigma^2_{\Sigma/\Sigma_0} / \sigma^2_{\rho/\rho_0}$ is only a function of $M_s$.

\begin{deluxetable*}{cccccc}
\tabletypesize{\scriptsize}
\tablecaption{Simulation Results}
\tablewidth{0pt}
\tablehead{ 
\colhead{Model} & 
\colhead{$M_{s}$} & 
\colhead{$N$ \tablenotemark{a}} & 
\colhead{$\sigma^{2}_{\rho / \rho_{0}}$ \tablenotemark{b}} & 
\colhead{$\sigma^{2}_{\Sigma_{x} / \Sigma_{x,0}}$ \tablenotemark{d}} &
\colhead{$\sigma^{2}_{\Sigma_{z} / \Sigma_{z,0}}$ \tablenotemark{e}} 
} 
\startdata
solF512-kf10b1  & $\sim$ 0.66 $\pm$ 0.01  & 1.25  & $\sim$ 0.07  $\pm$ 0.00  & $\sim$ 0.025 $\pm$ 0.00  & $\sim$ 0.018 $\pm$ 0.00 \\
                &                         & 2.5   & $\sim$ 0.07  $\pm$ 0.00  & $\sim$ 0.014 $\pm$ 0.00  & $\sim$ 0.010 $\pm$ 0.00 \\
                &                         & 5.0   & $\sim$ 0.07  $\pm$ 0.00  & $\sim$ 0.007 $\pm$ 0.00  & $\sim$ 0.005 $\pm$ 0.00 \\
                &                         & 10.0  & $\sim$ 0.07  $\pm$ 0.00  & $\sim$ 0.004 $\pm$ 0.00  & $\sim$ 0.003 $\pm$ 0.00 \\
\hline
                & $\sim$ 0.88 $\pm$ 0.01  & 1.25  & $\sim$ 0.16  $\pm$ 0.00  & $\sim$ 0.048 $\pm$ 0.00  & $\sim$ 0.038 $\pm$ 0.00 \\
                &                         & 2.5   & $\sim$ 0.16  $\pm$ 0.00  & $\sim$ 0.026 $\pm$ 0.00  & $\sim$ 0.020 $\pm$ 0.00 \\
                &                         & 5.0   & $\sim$ 0.16  $\pm$ 0.00  & $\sim$ 0.013 $\pm$ 0.00  & $\sim$ 0.011 $\pm$ 0.00 \\
                &                         & 10.0  & $\sim$ 0.16  $\pm$ 0.00  & $\sim$ 0.007 $\pm$ 0.00  & $\sim$ 0.006 $\pm$ 0.00 \\
\hline
                & $\sim$ 2.98  $\pm$ 0.02 & 1.25  & $\sim$ 1.28  $\pm$ 0.06  & $\sim$ 0.242 $\pm$ 0.01  & $\sim$ 0.248 $\pm$ 0.01 \\
                &                         & 2.5   & $\sim$ 1.28  $\pm$ 0.04  & $\sim$ 0.125 $\pm$ 0.00  & $\sim$ 0.132 $\pm$ 0.01 \\
                &                         & 5.0   & $\sim$ 1.28  $\pm$ 0.03  & $\sim$ 0.064 $\pm$ 0.00  & $\sim$ 0.068 $\pm$ 0.00 \\
                &                         & 10.0  & $\sim$ 1.28  $\pm$ 0.03  & $\sim$ 0.031 $\pm$ 0.00  & $\sim$ 0.035 $\pm$ 0.00 \\
\hline
                & $\sim$ 4.74 $\pm$ 0.02  & 1.25  & $\sim$ 1.90  $\pm$ 0.12  & $\sim$ 0.337 $\pm$ 0.02  & $\sim$ 0.338 $\pm$ 0.03 \\
                &                         & 2.5   & $\sim$ 1.90  $\pm$ 0.08  & $\sim$ 0.177 $\pm$ 0.01  & $\sim$ 0.180 $\pm$ 0.02 \\
                &                         & 5.0   & $\sim$ 1.90  $\pm$ 0.06  & $\sim$ 0.091 $\pm$ 0.00  & $\sim$ 0.092 $\pm$ 0.01 \\
                &                         & 10.0  & $\sim$ 1.90  $\pm$ 0.05  & $\sim$ 0.047 $\pm$ 0.00  & $\sim$ 0.047 $\pm$ 0.00 \\
\hline
                & $\sim$ 8.62 $\pm$ 0.01  & 1.25  & $\sim$ 2.82  $\pm$ 0.17  & $\sim$ 0.389 $\pm$ 0.03  & $\sim$ 0.438 $\pm$ 0.03 \\
                &                         & 2.5   & $\sim$ 2.82  $\pm$ 0.09  & $\sim$ 0.197 $\pm$ 0.01  & $\sim$ 0.230 $\pm$ 0.02 \\
                &                         & 5.0   & $\sim$ 2.82  $\pm$ 0.07  & $\sim$ 0.098 $\pm$ 0.01  & $\sim$ 0.118 $\pm$ 0.01 \\
                &                         & 10.0  & $\sim$ 2.82  $\pm$ 0.02  & $\sim$ 0.049 $\pm$ 0.00  & $\sim$ 0.061 $\pm$ 0.00 \\
\hline \hline
solF1024-kf10b1 & $\sim$ 3.16 $\pm$ 0.01  & 1.25  & $\sim$ 1.60  $\pm$ 0.09  & $\sim$ 0.275 $\pm$ 0.02  & $\sim$ 0.248 $\pm$ 0.02 \\
                &                         & 2.5   & $\sim$ 1.60  $\pm$ 0.08  & $\sim$ 0.141 $\pm$ 0.01  & $\sim$ 0.131 $\pm$ 0.01 \\
                &                         & 5.0   & $\sim$ 1.60  $\pm$ 0.07  & $\sim$ 0.072 $\pm$ 0.00  & $\sim$ 0.067 $\pm$ 0.00 \\
                &                         & 10.0  & $\sim$ 1.60  $\pm$ 0.07  & $\sim$ 0.038 $\pm$ 0.00  & $\sim$ 0.035 $\pm$ 0.00 \\
\hline
                & $\sim$ 10.40 $\pm$ 0.01 & 1.25  & $\sim$ 4.39  $\pm$ 0.24  & $\sim$ 0.471 $\pm$ 0.03  & $\sim$ 0.468 $\pm$ 0.03 \\
                &                         & 2.5   & $\sim$ 4.39  $\pm$ 0.18  & $\sim$ 0.243 $\pm$ 0.02  & $\sim$ 0.245 $\pm$ 0.01 \\
                &                         & 5.0   & $\sim$ 4.39  $\pm$ 0.09  & $\sim$ 0.123 $\pm$ 0.01  & $\sim$ 0.126 $\pm$ 0.01 \\
                &                         & 10.0  & $\sim$ 4.39  $\pm$ 0.06  & $\sim$ 0.063 $\pm$ 0.00  & $\sim$ 0.064 $\pm$ 0.00 \\
\hline \hline
solF512-kf2.5b1 & $\sim$ 0.81  $\pm$ 0.02 & 2.5   & $\sim$ 0.09  $\pm$ 0.01  & $\sim$ 0.016 $\pm$ 0.00  & $\sim$ 0.009 $\pm$ 0.00 \\
                & $\sim$ 3.40  $\pm$ 0.02 & 2.5   & $\sim$ 1.35  $\pm$ 0.28  & $\sim$ 0.136 $\pm$ 0.03  & $\sim$ 0.140 $\pm$ 0.02 \\
                & $\sim$ 5.24  $\pm$ 0.02 & 2.5   & $\sim$ 2.09  $\pm$ 0.25  & $\sim$ 0.190 $\pm$ 0.01  & $\sim$ 0.202 $\pm$ 0.03 \\
                & $\sim$ 6.87  $\pm$ 0.03 & 2.5   & $\sim$ 2.54  $\pm$ 0.33  & $\sim$ 0.198 $\pm$ 0.03  & $\sim$ 0.254 $\pm$ 0.05\\
\hline \hline
compF512-kf2.5b1 & $\sim$ 0.50  $\pm$ 0.02 & 2.5   & $\sim$ 0.27  $\pm$ 0.06  & $\sim$ 0.046 $\pm$ 0.02  & $\sim$ 0.050 $\pm$ 0.02 \\
                & $\sim$ 0.98  $\pm$ 0.02 & 2.5   & $\sim$ 0.73  $\pm$ 0.08  & $\sim$ 0.075 $\pm$ 0.03  & $\sim$ 0.162 $\pm$ 0.11 \\
                & $\sim$ 3.34  $\pm$ 0.02 & 2.5   & $\sim$ 6.01  $\pm$ 1.39  & $\sim$ 0.428 $\pm$ 0.12  & $\sim$ 0.658 $\pm$ 0.20 \\
                & $\sim$ 9.43  $\pm$ 0.01 & 2.5   & $\sim$ 25.11 $\pm$ 3.94  & $\sim$ 1.147 $\pm$ 0.16  & $\sim$ 1.528 $\pm$ 0.24     \\
\hline \hline              
solF512-kf2.5Hydro & $\sim$ 1.02  $\pm$ 0.02 & 2.5   & $\sim$ 0.10  $\pm$ 0.01  & $\sim$ 0.012 $\pm$ 0.00  & $\sim$ 0.013 $\pm$ 0.00 \\
                & $\sim$ 10.25 $\pm$ 0.01 & 2.5   & $\sim$ 6.39  $\pm$ 0.09  & $\sim$ 0.273 $\pm$ 0.05  & $\sim$ 0.255 $\pm$ 0.02   
\enddata
\tablecomments{
\tablenotetext{a}{The number of independent eddies along the line of sight.} 
\tablenotetext{b}{The variance of the 3D density normalized by the mean 3D density.}
\tablenotetext{d}{The variance  of column density normalized by the mean column density projected along the $x$-axis, which is parallel to the mean magnetic field $\bf{B}_{0}$. }
\tablenotetext{e}{The variance  of column density normalized by the mean column density projected along the $z$-axis, which is perpendicular to the mean magnetic field $\bf{B}_{0}$. }
}
\label{tbl:obs}
\end{deluxetable*}

\subsection{The relation between $ N \sigma^2_{\Sigma/\Sigma_0}$ and $\sigma^2_{\rho/\rho_0}$}

\begin{figure*}
\centering
\includegraphics[scale=.45]{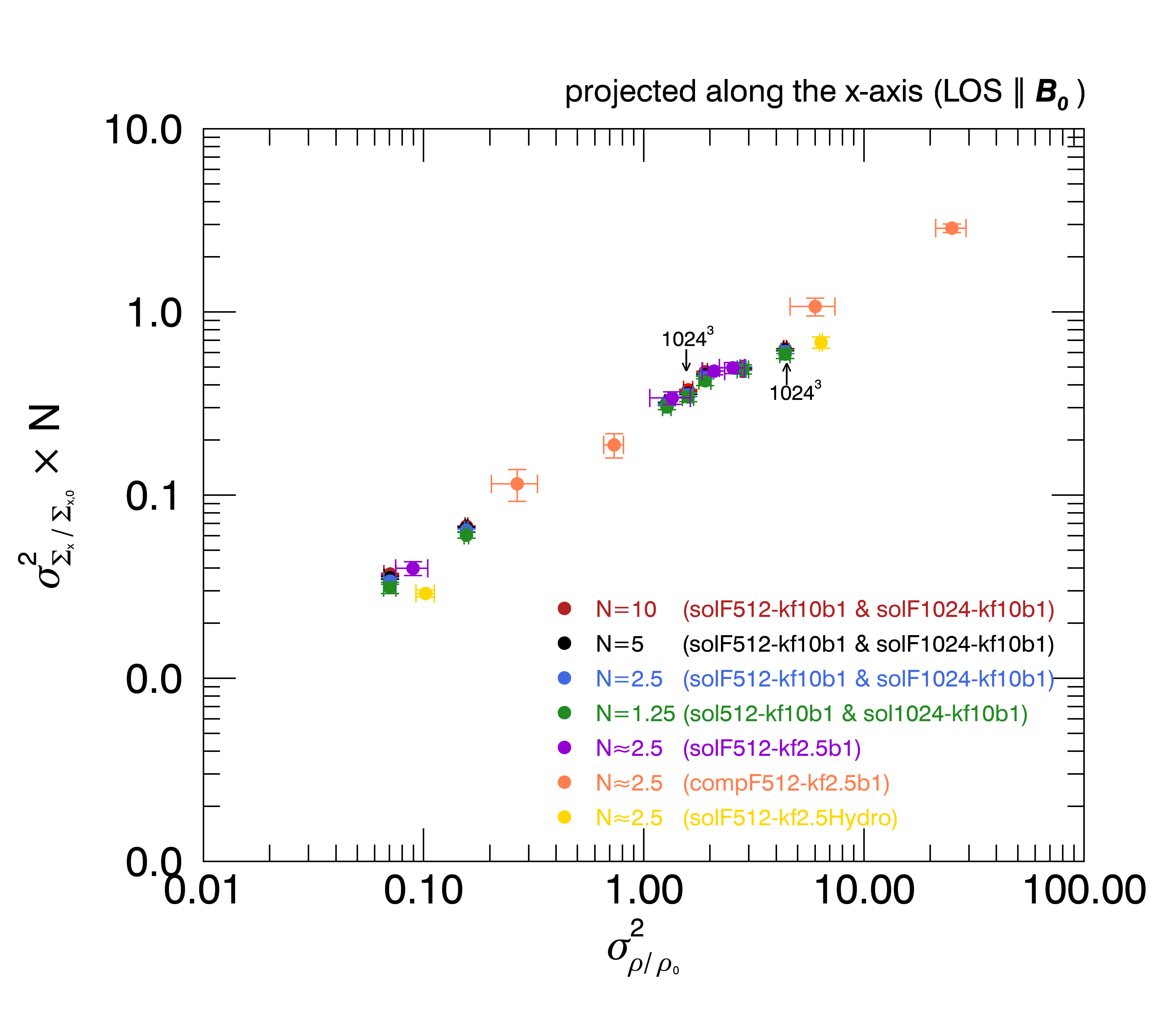}
\includegraphics[scale=.45]{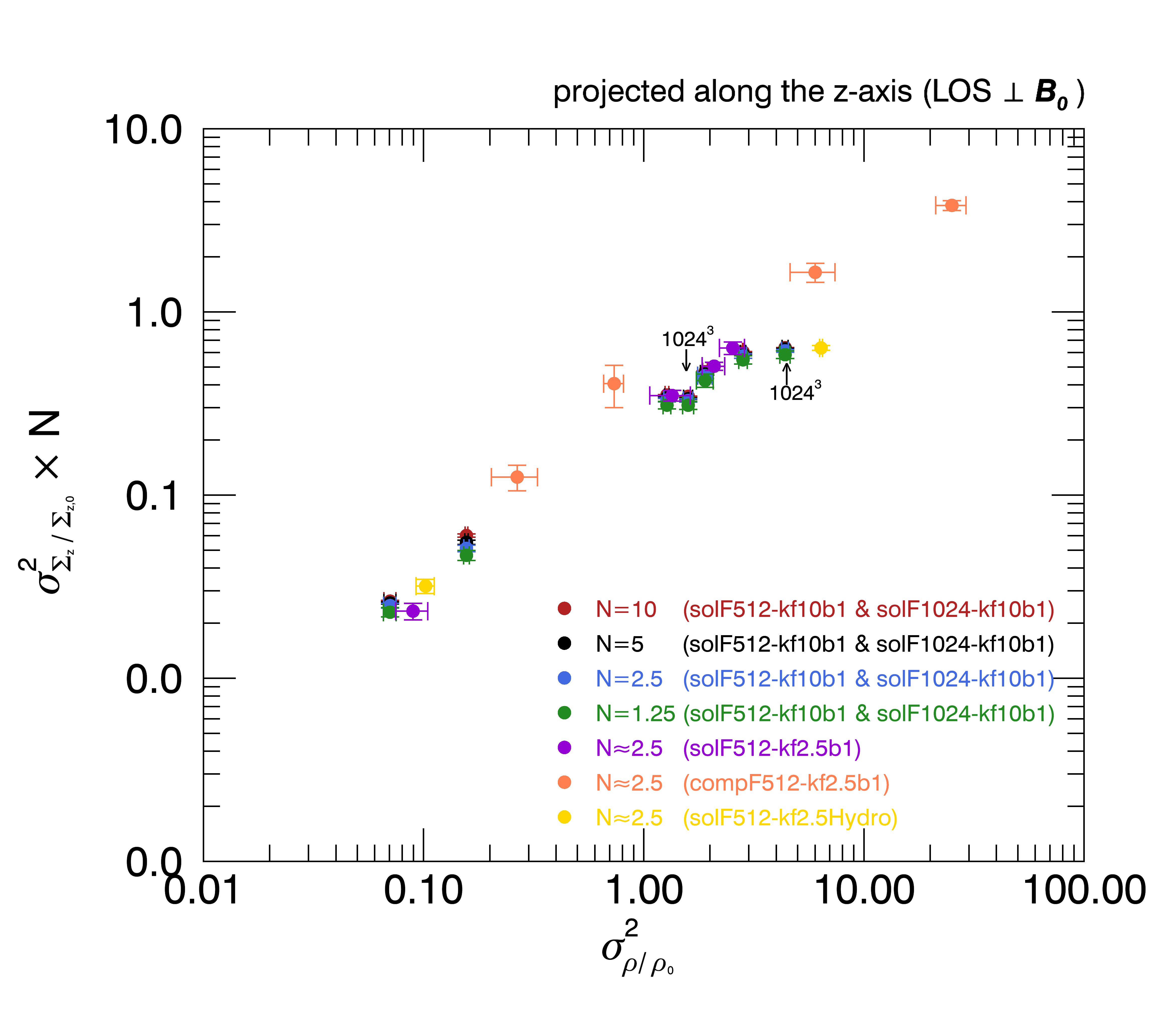}
\caption{Relation between $ N \sigma^2_{\Sigma/\Sigma_0}$ and $\sigma^2_{\rho/\rho_0}$. 
The line of sight is (a) along the $x$-axis and (b)  along the $z$-axis.
The colors of the circles correspond to those in Figure \ref{fig:R} 
 }
\label{fig:2D3D}
\end{figure*}

Figure \ref{fig:2D3D} shows the relation between $ N \sigma^2_{\Sigma/\Sigma_0}$ and $\sigma^{2}_{\rho / \rho_{0}}$.
As in Figure \ref{fig:R}, the left and the right panels are for the column density projected in directions parallel and
perpendicular to  the mean magnetic field, respectively.
As in Figure \ref{fig:R}, we use all the data listed  in Table \ref{tbl:obs}, which include both solenoidally and compressively driven turbulence.
As we can see in Figure \ref{fig:2D3D}, two quantities roughly lie on a single curve regardless of the sonic Mach number $M_s$.
Note that compressively driven MHD  turbulence and solenoidally driven hydrodynamic turbulence also follow the same relation.
Figure \ref{fig:2D3D} is useful to obtain the variance of 3D density  directly from 2D column density observations\footnote{
Note however  that the vertical axes contain $N$, the number of independent eddies along the line of sight.
Therefore, in order to obtain $\sigma^{2}_{\rho/\rho_0}$ directly from 
2D column density observation, we need to know $N$.}.
 
\section{Discussion} \label{sect:dis}

\subsection{Obtaining $\sigma_{\rho/\rho_0}$ from column density observations}
\label{sect:obtain3D}
From observations, we can obtain $\sigma_{\Sigma/\Sigma_0}$ 
 and spectrum of column density.
 From the latter we can calculate the Brunt factor $\mathcal{R}$.
 Then,
for a cubic cloud, we can easily obtain $\sigma_{\rho/\rho_0}$ from Equation (\ref{eq:cubeN}).
However, for a rectangular cloud, we need to know $L_\bot/L_{\rm los}$ or $N$ (see Equations (\ref{eq:rectR}) and  (\ref{eq:rectN})).

It may \emph{not} be possible to obtain  $L_\bot/L_{\rm los}$ or $N$ from column density spectrum.
The reason is as follows (see also discussions in Section \ref{sect:rect}). Consider Figure \ref{fig:shape}(a) and \ref{fig:shape}(b).
 If turbulence properties are similar in each cube, the shapes of the observed column density spectra (for $k \neq 0$) will be very similar for Figure \ref{fig:shape}(a) and \ref{fig:shape}(b). If the column density spectrum for Figure \ref{fig:shape}(a) is $Ak^{-m}$, that for Figure \ref{fig:shape}(b) will be  $\sim 2Ak^{-m}$. 
Therefore, the Brunt factor for both cases will be very similar and therefore we may not be able to derive information on $N$ from the column density spectrum.

From a separate observation, we may be able to constrain the value of $N$.
For example, if observation of an optically thin ratio emission line is available for many lines of sight, then we 
 may constrain $N$ from  the standard deviation  of the centroid velocity \citep[e.g.,][]{Cho2016}.

\subsection{Constraining $N$ from column density observations}
If we know $\sigma^2_{\Sigma/\Sigma_0}$ and $\sigma^2_{\rho/\rho_0}$, we can obtain $N$ 
(see Equation (\ref{eq:rectN})). 
In principle, we know $\sigma_{\rho/\rho_0}^2$ if we know the sonic Mach number $M_s$ and the mode of driving (i.e., solenoidal vs. compressive driving).
We can determine the sonic Mach number from the sound speed and the line-of-sight velocity dispersion of an optically thin radio emission line \citep[see for example discussions in][]{Stewart2022,Gerrard2023}.
Therefore, if we observe $\sigma_{\Sigma/\Sigma_0}^2$ and know $M_s$ and the mode of driving from separate observations, we can constrain the value of $N$.

\subsection{Effect of the size of the observed region on the sky} \label{sect:small}

\begin{figure*}
\centering
\includegraphics[scale=.2]{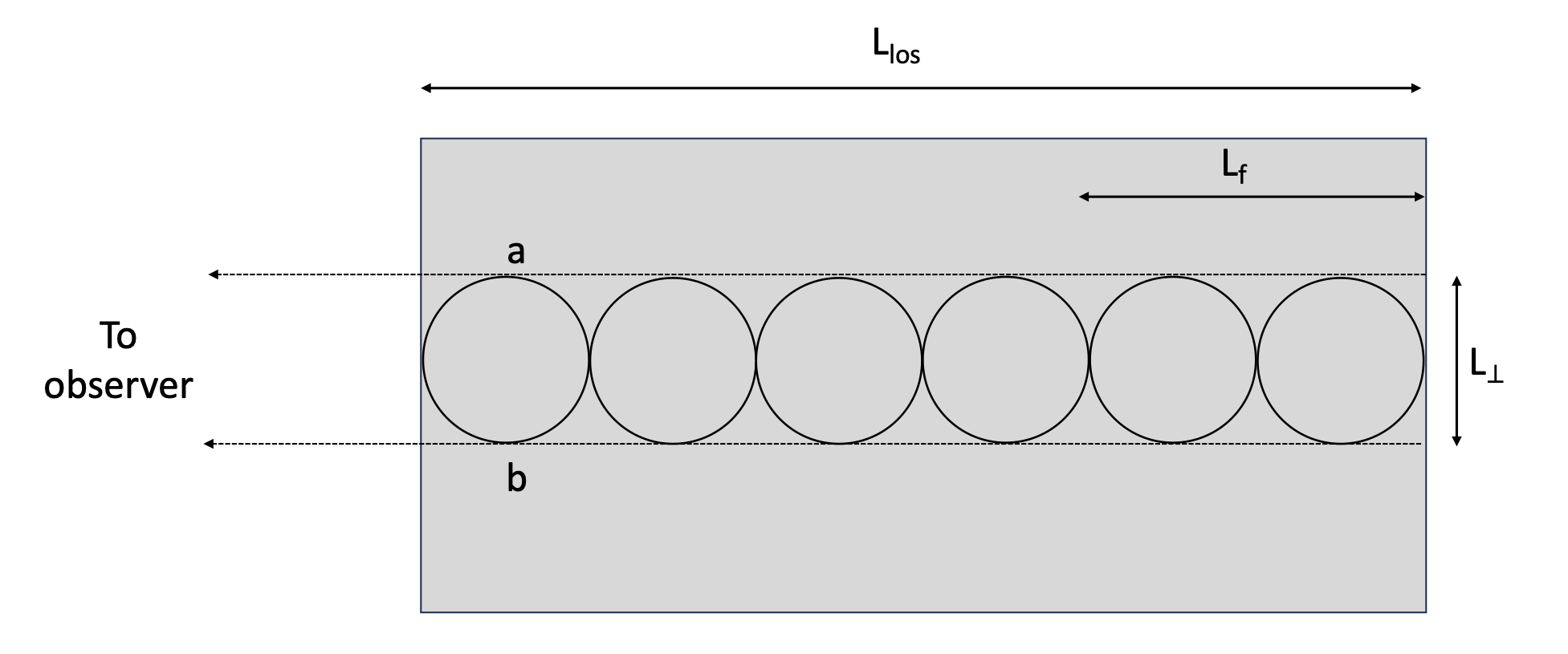}
\caption{Schematic explanation for the case of $L_\bot < L_f$. The physical size of the observed region projected on the sky is $L_\bot$. The 3D-density difference between points a and b, which are separated by $L_\bot$, will be $\sim \sigma_{\rho}(L_\bot/L_f)^{1/3}$ if we assume Kolmogorov spectrum for the 3D density.  }
\label{fig:small}
\end{figure*}

In this subsection, we discuss how $\sigma^2_{\Sigma/\Sigma_0} / \sigma^2_{\rho/\rho_0}$ changes as the physical size of the observed region on the sky ($L_\bot \times L_\bot$) changes.
Note that we assume $L_{\rm los}$ and $L_f$ are fixed in this subsection.
We consider two different cases: $L_\bot > L_f$ and $L_\bot < L_f$.

So far, we have assumed that $L_\bot > L_f$  (see the beginning of Section \ref{sect:theory}).
If $L_\bot > L_f$, we can use Equation (\ref{eq:rectN}):
\begin{equation}
\frac{ \sigma^2_{\Sigma/\Sigma_0} }{ \sigma^2_{\rho/\rho_0} }
        =\frac{ (3-m) }{2(2-m) } \frac{ L_f }{L_{\rm los} },
\end{equation}
which does not depend on $L_\bot$.
Therefore, $\sigma^2_{\Sigma/\Sigma_0} / \sigma^2_{\rho/\rho_0}$ remains constant when
the physical size of the observed region on the sky ($L_\bot \times L_\bot$) changes.

If $L_\bot < L_f$, we can show that  $\sigma^2_{\Sigma/\Sigma_0} / \sigma^2_{\rho/\rho_0}$ decreases
as the physical size of the observed region on the sky ($L_\bot \times L_\bot$) decreases.
 Suppose that physical distance between two lines of sight is $L_\bot$, which is smaller than the driving scale $L_f$ (see Figure \ref{fig:small}).
 Consider two points separated by $L_\bot$ (see, for example, two point a and b in Figure \ref{fig:small}).
The 3D-density difference between the points  will be
$\sim \sigma_{\rho / \rho_0}(L_\bot/L_f)^{1/3}$ if we assume Kolmogorov spectrum for the 3D density.
Therefore, the column density difference between two lines of sight separated by $L_\bot$ due to a single eddy of size $L_\bot$ 
will be $\sim \sigma_{\rho / \rho_{0}} L_\bot^{4/3}/L_f^{1/3}$.
There are $L_{\rm los}/L_\bot$ such eddies along the line of sight.
If they contribute stochastically, the column density difference between two lines of sight separated by $L_\bot$ due to
all eddies of size $L_\bot$ will be
\begin{equation}
\sigma_{\Sigma / \Sigma_{0}} \sim \sigma_{\rho /\rho_{0}} ( L_\bot^{4/3}/L_f^{1/3}) (L_{\rm los}/L_\bot)^{1/2},
\end{equation}
which results in
\begin{equation}
   \frac{ \sigma^2_{\Sigma/\Sigma_0} }{ \sigma^2_{\rho/\rho_0} } \sim \left(\frac{L_\bot}{L_f}\right)^{2/3} \frac{L_\bot }{L_{\rm los}}
   \propto L_\bot^{5/3}.
\end{equation}
Therefore, we expect  that  $\sigma^2_{\Sigma/\Sigma_0} / \sigma^2_{\rho/\rho_0}$ decreases
as $L_\bot$ decreases, which is seen in some observations \citep[see, for example,][]{Gerrard2023}.
In general, if the 3D power spectrum of density is proportional to $k^{-m}$ (with $m>3$), we can show that
\begin{equation}
\frac {\sigma^{2}_{\Sigma / \Sigma_{0}}} {\sigma^{2}_{\rho / \rho_{0}}}
\propto L_{\bot}^{m-2}.
\end{equation}
\\

\section{Conclusion}
\label{sect:summary}
In this paper, we have revisited the relation between the variance of the 3D density ($\sigma^{2}_{\rho / \rho_{0}}$) and that of the projected two-dimensional column density ($\sigma^{2}_{\Sigma / \Sigma_{0}}$) in turbulent media. 
Here,
$\sigma_{\Sigma/\Sigma_0}$ is the standard deviation of the column density normalized by
	the mean column density, and
	$\sigma_{\rho/\rho_0}$  is the standard deviation of the 3D density normalized by
	the mean 3D density.
In particular, we have investigated the behavior of
$\sigma^2_{\Sigma/\Sigma_0} / \sigma^2_{\rho/\rho_0}$. 
In this paper, we have mostly assumed that $L_\bot > L_f$ (see Section \ref{sect:small} for the case of $L_\bot < L_f$).
Our results show that the relative size of $L_{\rm los}$ and $L_\bot$ is important, where
$L_{\rm los}$ and $L_\bot$ are sizes of the cloud in
directions along and perpendicular to the line of sight, respectively.
We have found and numerically confirmed the following results.

\begin{itemize}
	\item  If $L_{\rm los} = L_\bot$,  
	$\sigma^2_{\Sigma/\Sigma_0} / \sigma^2_{\rho/\rho_0} = \mathcal{R}$, which is 
	identical to the earlier result by Brunt et al. (2010). Here, $\mathcal{R}$ is the Brunt factor.

	\item  If $L_{\rm los} \neq L_\bot$,  
	$\sigma^2_{\Sigma/\Sigma_0} / \sigma^2_{\rho/\rho_0} = \mathcal{R} L_\bot /L_{\rm los}$.
	
	\item For both $L_{\rm los}=L_\bot$ and $L_{\rm los}\neq L_\bot$ cases, $  \sigma^2_{\Sigma/\Sigma_0} / \sigma^2_{\rho/\rho_0} = \frac{ (3-m) }{2(2-m) } \frac{ 1}{N }$, 
 where $N$ is the number of independent eddies along the line of sight and  $-m$ is the power-law index of the 3D-density spectrum. Here we assume $m>3$  ($m=11/3$ for a Kolmogorov spectrum).

  \item When we plot $N\sigma^{2}_{\Sigma/\Sigma_0}$ against $\sigma^{2}_{\rho/\rho_0}$,  two quantities roughly lie on a single curve regardless of the sonic Mach number and the driving mode of turbulence.
\end{itemize}
In addition we have discussed observational implications.

\begin{acknowledgments}
This work is supported by the National R \& D Program
through the National Research Foundation of Korea Grants
funded by the Korean Government (NRF-2016R1D1A1B02015014).

\end{acknowledgments}


\end{document}